\documentclass[letterpaper,12pt]{article}

\usepackage{jheppub}

\usepackage{graphics,epsfig}
\usepackage{graphicx}

\usepackage{amsmath}
\usepackage{amssymb}
\usepackage{epsfig}

\def\half{\frac{1}{2}}

\usepackage{color}

\newcommand{\labell}[1]{\label{#1}}

\newcommand{\be}{\begin{equation}}
\newcommand{\ee}{\end{equation}}
\newcommand{\bea}{\begin{eqnarray}}
\newcommand{\eea}{\end{eqnarray}}
\newcommand{\ba}{\begin{eqnarray}}
\newcommand{\ea}{\end{eqnarray}}
\newcommand{\nn}{\nonumber \\}

\newcommand{\beq}{\begin{equation}}
\newcommand{\eeq}{\end{equation}}
\newcommand{\beqa}{\begin{eqnarray}}
\newcommand{\eeqa}{\end{eqnarray}}
\newcommand{\beqar}{\begin{eqnarray*}}
\newcommand{\eeqar}{\end{eqnarray*}}

\newcommand{\E}{\mathcal{E}}

\newcommand{\cO}{\mathcal{O}}

\newcommand{\la}{\lambda}

\newcommand\ra{\rangle}
\renewcommand\la{\langle}

\newcommand{\tr}{{\rm tr}}
\newcommand{\Tr}{{\rm Tr}}

\def\({\left(} \def\){\right)}
\def\[{\left[} \def\]{\right]}

\newcommand{\lie}[1]{ \pounds_{#1}}

\title{Entanglement entropy from one-point functions in holographic states}

\author[]{Matthew J.~S.~Beach,}
\author[]{Jaehoon Lee,}
\author[]{Charles Rabideau,}
\author[]{Mark Van Raamsdonk}

\affiliation[]{Department of Physics and Astronomy, University of British Columbia \\ 6224 Agricultural Road, Vancouver, BC, V6T 1W9, Canada}

\emailAdd{mbeach,jaehlee,rabideau,mav@phas.ubc.ca}

\abstract
{
For holographic CFT states near the vacuum, entanglement entropies for spatial subsystems can be expressed perturbatively as an expansion in the one-point functions of local operators dual to light bulk fields. Using the connection between quantum Fisher information for CFT states and canonical energy for the dual spacetimes, we describe a general formula for this expansion up to second-order in the one-point functions, for an arbitrary ball-shaped region, extending the first-order result given by the entanglement first law. For two-dimensional CFTs, we use this to derive a completely explicit formula for the second-order contribution to the entanglement entropy from the stress tensor. We show that this stress tensor formula can be reproduced by a direct CFT calculation for states related to the vacuum by a local conformal transformation. This result can also be reproduced via the perturbative solution to a non-linear scalar wave equation on an auxiliary de Sitter spacetime, extending the first-order result in arXiv/1509.00113.
}

\begin{document}

\maketitle

\section{Introduction}

In holographic conformal field theories, states with a simple classical gravity dual interpretation have a remarkable structure of entanglement: according to the holographic entanglement entropy formula~\cite{Ryu:2006bv, Ryu:2006ef, Hubeny:2007xt}, their entanglement entropies for arbitrary regions (at leading order in large $N$) are completely encoded in the extremal surface areas of an asymptotically AdS spacetime. In general, the space of possible entanglement entropies (functions on a space of subsets of the AdS boundary) is far larger than the space of possible asymptotically AdS metrics (functions of a few spacetime coordinates), so this property of geometrically-encodable entanglement entropy should be present in only a tiny fraction of all quantum field theory states~\cite{Lashkari:2014kda}. It is an interesting question to understand better which CFT states have this property\footnote{Even in holographic CFTs, it is clear that not all states will have this property. For example, if $|\Psi_1 \rangle$ and $|\Psi_2 \rangle$ are two such states, corresponding to different spacetimes $M_{\Psi_1}$ and $M_{\Psi_2}$, the superposition $|\Psi_1 \rangle + |\Psi_2 \rangle$ is not expected to correspond to any single classical spacetime but rather to a superposition of $M_{\Psi_1}$ and $M_{\Psi_2}$. Thus, the set of ``holographic states'' is not a subspace, but some general subset.}, and which properties of a CFT will guarantee that families of low-energy states with geometric entanglement exist.

For a hint towards characterizing these holographic states, consider the gravity perspective.  A spacetime $M_\Psi$ dual to a holographic state $|\Psi \rangle$ is a solution to the bulk equations of motion. Such a solution can be characterized by a set of initial data on a bulk Cauchy surface (and appropriate boundary conditions at the AdS boundary). The solution away from the Cauchy surface is determined by evolving this initial data forwards (or backwards) in time using the bulk equations. Alternatively, we can think of the bulk solution as being determined by evolution in the holographic radial direction, with ``initial data'' specified at the timelike boundary of AdS. In this case, the existence and uniqueness of a solution is more subtle, but the asymptotic behavior of the fields determines the metric at least in a perturbative sense (e.g. perturbatively in deviations from pure AdS, or order-by-order in the Fefferman-Graham expansion). It is plausible that in many cases, this boundary data is enough to determine a solution nonperturbatively to some finite distance into the bulk, or even for the whole spacetime. Thus, for geometries dual to holographic states, we can say that the bulk spacetime (at least in a perturbative sense) is encoded in the boundary behavior of the various fields.

According to the AdS/CFT dictionary, this boundary behavior is determined by the one-point functions of low-dimension local operators associated with the light bulk fields. On the other hand, the bulk spacetime itself allows us to calculated entanglement entropies (and many other non-local quantities). Thus, the assumption that a state is holographic allows us (via gravity calculations) to determine the entanglement entropies and other non-local properties of the state (again, at least perturbatively) from the local data provided by the one-point functions:
\be
\label{Gravrules}
|\Psi \rangle \rightarrow \langle {\cal O}_\alpha(x^\mu)\rangle \rightarrow \phi_\alpha {\rm \; asymptotics}  \rightarrow \phi_\alpha (x^\mu, z) \rightarrow {\rm entanglement \; entropies \;} S(A)
\ee
where $\phi$ here indicates all light fields including the metric.\footnote{Here, the region $A$ should be small enough so that the bulk extremal surface associated with $A$ should be contained in the part of the spacetime determined through the equations of motion by the boundary values; we do not need this restriction if we are working perturbatively.}$^,\,$\footnote{Results along these lines in the limit of small boundary regions or constant one-point functions appeared in~\cite{Blanco:2013joa, Banerjee:2014oaa, Banerjee:2014ozp, Giusto:2014aba, Giusto:2015dfa}.}

The recipe (\ref{Gravrules}) could be applied to any state, but for states that are not holographic, the results will be inconsistent with the actual CFT answers. Thus, we have a stringent test for whether a CFT state has a dual description well-described by a classical spacetime: carry out the procedure in (\ref{Gravrules})
and compare the results with a direct CFT calculation of the entanglement entropies; if there is a mismatch for any region, the state is not holographic.\footnote{Another interesting possibility is that the one-point functions could give boundary data that is not consistent with any solution of the classical bulk equations; this possibility exists since the ``initial data'' for the radial evolution problem obeys certain constraints.}

In this paper, our goal is to present some more explicit results for the gravity prediction $S_A^{grav}(\langle {\cal O}_\alpha \rangle)$ in cases where the gravitational equations are Einstein gravity with matter and the region is taken to be a ball-shaped region $B$. We will work perturbatively around the vacuum state to obtain an expression as a power series in the one-point functions of CFT operators. At first-order, the result depends only on the CFT stress tensor expectation value \cite{Casini:2011kv}:
\be
\label{firstorder}
S_B(|\Psi \rangle) =  S^{vac}_B +  2 \pi \int_B d^{d-1} x {R^2 - r^2 \over 2 R} \langle T_{00} \rangle + {\cal O}(\langle {\cal O}_\alpha \rangle^2) \,.
\ee
This well-known expression is universal for all CFTs since it follows from the first law of entanglement $\delta^{(1)} S_B = \delta \langle H_B \rangle$, where
\be
\label{modH}
H_B \equiv -\log \rho_B^{vac} = 2 \pi \int_B d^{d-1} x {R^2 - r^2 \over 2 R} T_{00}
\ee
is the vacuum modular Hamiltonian for a ball-shaped region. Thus, to first-order, the gravity procedure (\ref{Gravrules}) always gives the correct CFT result for ball-shaped regions, regardless of whether the state is holographic.

\subsubsection*{General second-order result for ball entanglement entropy}

Our focus will be on the second-order answer; in this case, it is less clear whether the gravity results from (\ref{Gravrules}) should hold for any CFT or whether they represent a constraint from holography. To obtain explicit formulae at this order, we begin by writing
\be
S_B(|\Psi \rangle) = S_B^{vac} + \Delta \langle H_B \rangle - S(\rho_B || \rho_B^{vac})
\ee
which follows immediately from the definition of relative entropy $S(\rho_B || \rho_B^{vac})$ reviewed in Section \ref{Section:background} below. We then make use of a recent result in \cite{Lashkari:2015hha}: to second-order in perturbations from the vacuum state, the relative entropy for a ball-shaped region in a holographic state\footnote{This second-order relative entropy is known as quantum Fisher information.} is equal to the ``canonical energy'' associated with a corresponding wedge of the bulk spacetime. We provide a brief review of this in Section \ref{Section:background} below. On shell, the latter quantity can be expressed as a quadratic form on the space of first-order perturbations to pure AdS spacetime, so we have
\begin{align}
S(\rho_B||\rho_B^{vac}) = \Delta \langle H_B \rangle  - \Delta S_B  = \half {\cal E}(\delta \phi_\alpha, \delta \phi_\alpha) + {\cal O}(\delta \phi^3) \,.
\end{align}
Rearranging this, we have a second-order version of (\ref{firstorder}):
\begin{align}
\label{secondorder1}
S_B(|\Psi \rangle)
&= S_B^{vac} + \delta^{(1)}S_B + \delta^{(2)}S_B + \mathcal{O}(\delta \phi^3) \nn
&= S^{vac}_B + \Delta \langle H_B \rangle -  \half {\cal E}(\delta \phi_\alpha, \delta \phi_\alpha) + {\cal O}(\delta \phi^3) \nn
&= S^{vac}_B +  2 \pi \int_B d^{d-1} x {R^2 - r^2 \over 2 R} \langle T_{00} \rangle - \half {\cal E}(\delta \phi_\alpha, \delta \phi_\alpha) + {\cal O}(\delta \phi^3) \,.
\end{align}
As we review in Section~\ref{Section:background} below, the last term can be written more explicitly as
\be
\label{secondorder2}
{\cal E}(\delta \phi_\alpha, \delta \phi_\alpha) = \int_\Sigma \omega(\delta g, \lie{\xi} \delta g) -  \int_\Sigma \xi^a T^{(2)}_{ab}\epsilon^b\;,
\ee
where $\Sigma$ is a bulk spatial region between $B$ and the bulk extremal surface $\tilde{B}$ with the same boundary, $\omega$ is the ``presymplectic form'' whose integral defines the symplectic form on gravitational phase space,  $T^{(2)}_{ab}$ is the matter stress tensor at second-order in the bulk matter fields, and $\xi$ is a bulk Killing vector which vanishes on $\tilde{B}$. The first-order bulk perturbations  $\delta \phi_\alpha$ (including the metric perturbation) may be expressed in terms of the boundary one-point functions via bulk-to-boundary propagators
\be
\label{propagators}
\delta \phi_\alpha(x, z) = \int_{D_B} K_\alpha( x, z ;x') \langle {\cal O}_\alpha (x') \rangle \,,
\ee
where $D_B$ is the domain of dependence of the ball $B$.  Given the one-point functions within $D_B$, we can use \eqref{propagators} to determine the linearized bulk perturbation in $\Sigma$ and evaluate \eqref{secondorder2}.

The expression (\ref{secondorder1}), (\ref{secondorder2}), and (\ref{propagators}) together provide a formal result for the ball entanglement entropy of a holographic state, expanded to second-order in the boundary one-point functions.

\subsubsection*{Explicit results for 1+1 dimensional CFTs}

In order to check the general formula and provide more explicit results, we focus in Section \ref{Sec:Gravity} on the case of 1+1 dimensional CFTs, carrying out an explicit calculation of the gravitational contributions to (\ref{secondorder2}) starting from a general boundary stress tensor. We find the result
\be
\label{result2DS2}
 \delta^{(2)}S_B^{grav} = -\half \int_{B'} d x^+_1 \, \int_{B'} d x^+_2 \, K_2 (x^+_1,  x^+_2) \langle T_{++} (x^+_1) \rangle \langle T_{++} (x^+_2) \rangle + \left\{+ \leftrightarrow-\right\}
\ee
where the integrals can be taken over any spatial surface $B'$ with boundary $\partial B$, and the kernel is given by
\be
\label{result2D}
K_2 (x_1,  x_2) = \frac{6 \pi ^2 }{c R^2}  \left\{ \begin{array}{ll}
 (R-x_1)^2 (R+x_2)^2 & \quad x_1\geq x_2 \\
 (R+x_1)^2 (R-x_2)^2 & \quad x_1 < x_2
\end{array} \right. \, ,
\ee
where $c$ is the central charge. In this special case, the conservation equations determine the stress tensor expectation values throughout the region $D_B$ from the expectation values on $B'$, so as in the first-order result (\ref{firstorder}), our final expression involves integrals only over $B'$. This will not be the case for the terms involving matter fields, or in higher dimensions. As a consistency check, we show that the expression (\ref{result2D}) is always negative, as required by its interpretation as the second-order contribution to relative entropy.

We can also check the formula (\ref{result2D}) via a direct CFT calculation by considering states that are obtained from the CFT vacuum by a local conformal transformation. In two dimensions, states with an arbitrary traceless conserved stress-tensor can be obtained, and the entanglement entropy for these states can also be calculated explicitly. We carry out this calculation in section 4, and show that the result (\ref{result2D}) is exactly reproduced.

In Section~\ref{sec:scalar}, we consider the matter terms in (\ref{secondorder2}) providing some explicit results for the quadratic contributions of scalar operator expectation values. Here, as in the generic case, the result takes the form
\be
\delta^{(2)}S_B^{matter} = -\half \int_{D_B} \int_{D_B} G_{\alpha \beta}(x,x') \langle {\cal O}_\alpha(x) \rangle \langle {\cal O}_\beta(x') \rangle
\ee
with integrals over the entire domain of dependence region.

\subsubsection*{Auxiliary de Sitter Space Interpretation}

Recently, in~\cite{deBoer:2015kda} it has been pointed out that the first-order result $\delta^{(1)} S(x^\mu,R)$ for the entanglement entropy of a ball with radius $R$ and center $x^\mu$ can be obtained as the solution to the equation of motion for a free scalar field on an auxiliary de Sitter space $ds^2 = \tfrac{L_{dS}^2}{R^2}(-dR^2 + dx_\mu dx^\mu)$ with the CFT energy density $\langle T_{00}(x^\mu) \rangle$ acting as a source term at $R=0$. In Section~\ref{Sec:EntanglementHolography}, we show that in the 1+1 dimensional case, the stress tensor term (\ref{result2D}) for the entanglement entropy at second-order can also results from solving a scalar field equation on the auxiliary de Sitter space if we add a simple cubic interaction term. In an upcoming paper \cite{helleretal}, it is shown that this agreement extends to all orders for a suitable choice of the scalar field potential. The resulting nonlinear wave equation also reproduces the second-order entanglement entropy near a thermal state in the auxiliary kinematic space recently described in~\cite{Asplund:2016koz}.

Including the contributions from matter fields or moving to higher dimensions, the expression for entanglement entropy involves one-point functions on the entire causal diamond $D_B$, so reproducing these results via some local differential equation will require a more complicated auxiliary space that takes into account the time directions in the CFT. This direction is pursued further in \cite{Czech:2016xec,helleretal}.

\subsubsection*{Discussion}

While the explicit two-dimensional stress tensor contribution (\ref{result2D}) can be obtained by a direct CFT calculation for a special class of states, we emphasize that in general the holographic predictions from (\ref{Gravrules}) are expected to hold only for holographic states in CFTs with gravity duals. It would be interesting to understand better whether all of the second order contributions we considered here are universal for all CFTs or whether they represent genuine constraints/predictions from holography.\footnote{There is evidence in~\cite{Faulkner:2014jva, Speranza:2016jwt, Faulkner:2015csl} that at least some of the contributions at this order can be reproduced by CFT calculations in general dimensions, since they arise from CFT two and three-point functions, though the results there most directly apply to the case where the perturbation is to the theory rather than the state.} In the latter case, and for the results at higher order in perturbation theory, it is an interesting question to understand better which CFT states and/or which CFT properties are required to reproduce the results through direct CFT calculations. This should help us understand better which theories and which states in these theories are holographic.


\section{Background}
\label{Section:background}
Our holographic calculation of entanglement entropy to second-order in the boundary one-point functions makes use of the direct connection between CFT quantum Fisher information and canonical energy on the gravity side, pointed out recently in \cite{Lashkari:2015hha}. We begin with a brief review of these results.

\subsection{Relative entropy and quantum Fisher information}

Our focus will be on ball-shaped subsystems $B$ of the CFT$_{d}$, for which the the vacuum density matrix is known explicitly through (\ref{modH}). More generally, we can write it as
\be
\label{modH2}
\rho_B^{vac} = e^{-H_B}\,, \qquad \qquad  H_B = \int_{B'} \zeta_B^{\mu} T_{\mu \nu} \epsilon^\nu\,,
\ee
where $T_{\mu \nu}$ is the CFT stress tensor operator and $\epsilon$ is defined as
\be
\label{defeps}
\epsilon_\nu = {1 \over (d-1)!} \epsilon_{\nu \nu_1 \cdots \nu_{d-1}} dx^{\nu_1} \wedge \cdots \wedge dx^{\nu_{d-1}}\,,
\ee
so that $n^\mu \epsilon_\mu$ is the volume form on the surface perpendicular to a unit vector $n^\mu$, and $\zeta_B$ is a conformal Killing vector defined in the domain of dependence region $D_B$, with $\zeta_B = 0$ on $\partial B$. For the ball $B$ with radius $R$ and center $x_0^\mu$ in the $t=t_0$ slice, we have
\be\labell{defzeta}
\zeta_B  = - \frac{2\pi}{ R}  (t-t_0) (x^i-x^i_0) \partial_i + \frac{\pi}{R} [R^2 - (t-t_0)^2 - (\vec{x}-\vec{x_0})^2] \, \partial_t \; .
\ee
By the conservation of the current $\zeta_B^\mu T_{\mu} {}^{\nu}$ associated with this conformal Killing vector, the integral in (\ref{modH2}) can be taken over any spatial surface $B'$ in $D_B$ with the same boundary as $B$.

For excited states, the density matrix $\rho_B$ will generally be different than $\rho_B^{vac}$. One measure of this difference is the relative entropy
 \bea
S(\rho_B || \rho^{vac}_B) &=& \textrm{tr}(\rho_B \log \rho_B)-\textrm{tr}(\rho_B \log \rho^{vac}_B) \cr
&=& \Delta \langle H_B \rangle -\Delta S_B \, ,
\label{REdef}
 \eea
where $H_B$ is the vacuum modular Hamiltonian given in (\ref{modH2}), $S_B = -\tr(\rho_B \log \rho_B)$ is the entanglement entropy for the region $B$ and $\Delta$ indicates the difference with the vacuum state.

For a one-parameter family of states near the vacuum, we can expand $\rho_B$ as
 \be
 \rho_B (\lambda) = \rho^{vac}_B+ \lambda \; \delta \rho_1 + \lambda^2 \delta \rho_2 + \cO(\lambda^3)\,.
 \ee
The  first-order contribution to relative entropy vanishes (this is the first law of entanglement $\delta^{(1)} S_B=\delta \la H_B\ra$) so the leading contribution to relative entropy appears at second-order in $\lambda$. This quadratic in $\delta \rho_1$ with no contribution from $\delta \rho_2$,
\be
\label{REpert}
S(\rho_B(\lambda) || \rho^{vac}_B ) = \lambda^2 \, \langle \delta \rho_1, \delta \rho_1 \rangle_{\rho^{vac}_B} + \cO(\lambda^3)\,,
\ee
where
\be
\langle \delta \rho, \delta \rho \rangle_\sigma \equiv \half \tr \left( \delta \rho \frac{d}{d \lambda} \log(\sigma + \lambda  \delta \rho) \Big|_{\lambda=0} \right) \; .
\ee
This quadratic form, which is positive by virtue of the positivity of relative entropy, defines a positive-(semi)definite metric on the space of perturbations to a general density matrix $\sigma$. This is known as the quantum Fisher information metric.

Rearranging (\ref{REdef}) and making use of (\ref{REpert}), we have
\be
\label{Sexp}
S_B = S_B^{vac} + \int_{B'} \zeta_B^{\mu}  \langle T_{\mu \nu} \rangle \epsilon^\nu - \lambda^2 \langle \delta \rho_1, \delta \rho_1 \rangle_{\rho^{vac}_B} + \cO(\lambda^3) \; .
\ee
This general expression is valid for any CFT, but the ${\cal O}(\lambda^2)$ term generally has no simple expression in terms of local operator expectation values. However, for holographic states we can convert this term into an expression quadratic in the CFT one-point functions by using the connection between quantum Fisher information and canonical energy.

\subsection{Canonical energy}

Consider now a holographic state, which by definition is associated with some dual asymptotically AdS spacetime $M$. Near the boundary, we can describe $M$ using a metric in Fefferman-Graham coordinates as
\be
\label{FG}
ds^2 = {\ell_{AdS}^2 \over z^2} \left( dz^2 + dx_\mu dx^\mu +  z^d \, \Gamma_{\mu \nu}(x, z) dx^\mu dx^\nu \right)
\ee
where $\Gamma_{\mu \nu}(z,x)$ has a finite limit as $z \to 0$ and $\Gamma=0$ for pure AdS.

The relative entropy $S(\rho_B || \rho_B^{vac})$ can be computed at leading order in large $N$ by making use of the holographic entanglement entropy formula, which relates the entanglement entropy for a region $A$ to the area of the minimal-area extremal surface $\tilde{A}$ in $M$ with boundary $\partial A$,
\begin{align}
\label{eq:rel-ent}
	 S_A & \equiv \frac{ {\rm Area} (\tilde{A})}{4\,G_N}  \, .
\end{align}
This yields immediately that  $\Delta S_A = ({\rm Area}(\tilde A)_M - {\rm Area}(\tilde A)_{AdS})/(4 G_N)$. The result (\ref{eq:rel-ent}) also allows us to relate the $\Delta \langle H_B \rangle$ term in relative entropy to a gravitational quantity, since it implies that the expectation value of the CFT stress tensor is related to the asymptotic behaviour of the metric through~\cite{Faulkner:2013ica}
\begin{align}
	\la T_{\mu\nu} \ra &= \frac{d \ell_{AdS}^{d-1}}{16 \pi G_N} \Gamma_{\mu\nu}(x, z=0) \,.
\label{eq:HoloRenT}
\end{align}
Thus, for holographic states, we can write
\be
\label{REgrav}
S(\rho_B||\rho_B^{vac}) = \frac{d \ell_{AdS}^{d-1}}{16 \pi G_N}  \int_{B} \zeta_B^{\mu} \Gamma_{\mu\nu}(x, 0)\, \epsilon^\nu - \frac{{\rm Area}(\tilde A)_M - {\rm Area}(\tilde A)_{AdS}}{4 G_N} \,.
\ee

For a one-parameter family of holographic states $|\Psi(\lambda) \rangle$ near the CFT vacuum, the dual spacetimes $M(\lambda)$ can be described via a metric and matter fields $\phi_\alpha = (g, \phi^{matter})$ with some perturbative expansion
\begin{align}
g &= g_{AdS} + \lambda \delta g_1 + \lambda^2 \delta g_2 + {\cal O} (\lambda^3)\;,   \nn
 \phi^{matter} &= \lambda \delta \phi^{matter}_1 + \lambda^2 \delta \phi^{matter}_2 + {\cal O} (\lambda^3) \;.
\end{align}
By the result (\ref{Sexp}) from the previous section, the second-order contribution to entanglement entropy is equal to the leading order contribution to relative entropy. This is related to a gravitational quantity via (\ref{REgrav}). The main result in \cite{Lashkari:2015hha} is that this second-order quantity can be expressed directly as a bulk integral over the spatial region $\Sigma$ between $B$ and $\tilde{B}$ where the integrand is a quadratic form on the linearized bulk perturbations $\delta g_1 $ and $\delta \phi^{matter}_1$.

\begin{figure}[h!]
  \centering
  \includegraphics{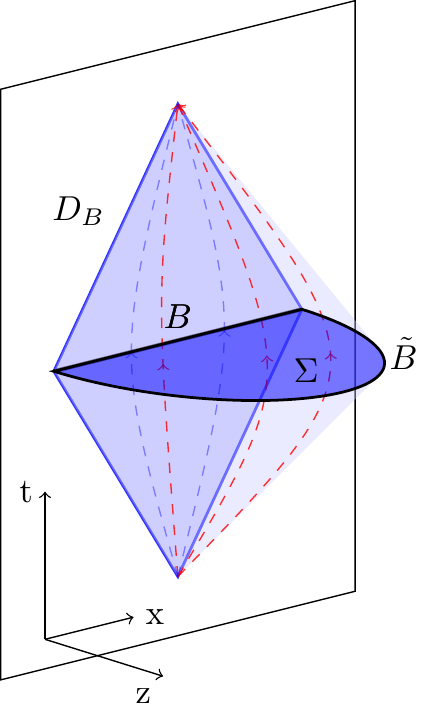}
  \caption{The Rindler wedge $R_B$ associated to the ball-shaped region $B$ on the boundary. The blue lines indicate the flow of $\zeta_B$, and the red lines $\xi_B$. The surface $\Sigma$ lies between $B$ and the extremal surface $\tilde B$.}\label{fig:rindler-wedge}
\end{figure}

To describe the general result, consider the region $\Sigma$ between $B$ and $\tilde{B}$ in pure AdS spacetime, and define $R_B$ as the domain of dependence of this region, as shown in figure \ref{fig:rindler-wedge}. Alternatively, $R_B$ is the intersection of the causal past and the causal future of $D_B$; it can be thought of as a Rindler wedge of AdS associated with $B$. On $R_B$, there exists a Killing vector which vanishes at $\tilde{B}$ and approaches the conformal Killing vector $\zeta_B$ at the boundary. In Fefferman-Graham coordinates, this is
\be
\label{defxi}
\xi_B  = - \frac{2\pi}{ R}  (t-t_0) [z \partial_z + (x^i-x^i_0) \partial_i ] + \frac{\pi}{R} [R^2 - z^2 - (t-t_0)^2 - (\vec{x}-\vec{x_0})^2] \, \partial_t
\ee
The vector $\xi_B$ is timelike hence defines a notion of time evolution within the region $R_B$; the ``Rindler time'' associated with this Rindler wedge.

The ``canonical energy'', dual to relative entropy at second-order, can be understood as the perturbative energy associated with this time, as explained in \cite{Hollands:2012sf}. This is quadratic in the perturbative bulk fields including the graviton, and given explicitly by
\bea
\label{canEn}
{\cal E}(\delta g_1, \delta \phi_1)   &=& W_\Sigma \left(\delta \phi_1 , \lie{\xi_B} \delta \phi_1 \right) \cr
&=&  \int_\Sigma \omega^{full} \left( \delta \phi_1, \lie{\xi_B} \delta \phi_1 \right) \cr
&=&  \int_\Sigma \omega \left( \delta g_1, \lie{\xi_B} \delta g_1\right) + \int_\Sigma \omega^{matter} \left( \delta \phi_1, \lie{\xi_B} \delta \phi_1 \right) \cr
&=&  \int_\Sigma \omega(\delta g_1, \lie{\xi_B} \delta g_1) - \int_\Sigma \xi^a_B T^{(2)}_{ab} \epsilon^b \; .
\eea
In the first line, $W_\Sigma$ is the symplectic form associated with the phase space of gravitational solutions on $\Sigma$, and $\lie{\xi_B} \delta \phi_1$ is the Lie derivative with respect to $\xi$ on $\delta \phi_1$,  the first-order perturbation in metric and matter fields. The symplectic form is equal to the integral over $\Sigma$ of a ``presymplectic'' form $\omega^{full}$ which splits into a gravitational part and a matter part as in the third line. The matter part can be written explicitly in terms of   $T^{(2)}_{ab}$, the matter stress tensor at quadratic order in the fields, while the gravitational part $\omega$ is given explicitly by
\bea
\label{defomega2}
\omega(\gamma^1, \gamma^2) &=& {1 \over 16 \pi G_N} \epsilon_a P^{abcdef} (\gamma^2_{bc} \nabla_d \gamma^1_{ef} - \gamma^1_{bc} \nabla_d \gamma^2_{ef}) \\
P^{abcdef} &=& g^{ae} g^{fb} g^{cd} - {1 \over 2} g^{ad} g^{be} g^{fc} - {1 \over 2} g^{ab} g^{cd} g^{ef} - {1 \over 2} g^{bc} g^{ae} g^{fd} + {1 \over 2} g^{bc} g^{ad} g^{ef} \; . \nonumber
\eea
In deriving (\ref{canEn}) it has been assumed that the metric perturbation has been expressed in a gauge for which the coordinate location of the extremal surface $\tilde{B}$ does not change (so that $\xi_B$ continues to vanish there), and the vector $\xi_B$ continues to satisfy the Killing equation at $\tilde{B}$. Thus, we require that
\bea
\label{eq:gauge-condition}
\xi_B |_{\tilde B (\lambda)} &= 0,\\
\label{eq:gauge-condition2}
\lie{\xi_B} g(\lambda) |_{\tilde B (\lambda)} &= 0.
\eea
As shown in \cite{Hollands:2012sf}, it is always possible to satisfy these conditions; we will see an explicit example below.

\section{Second-order contribution to entanglement entropy}
\label{Sec:Gravity}
Using the result (\ref{secondorder2}), we can now write down a general expression for the ball entanglement entropy of a general holographic state up to second-order in perturbations to the vacuum state, in terms of the CFT one-point functions. According to (\ref{Sexp}) and (\ref{canEn}), the second-order term in the entanglement entropy for a ball $B$ can be expressed as an integral over the bulk spatial region $\Sigma$ between $B$ and the corresponding extremal surface $\tilde{B}$, where the integrand is quadratic in first-order bulk perturbations.

These linearized perturbations are determined by the boundary behavior of the fields via the linearized bulk equations. In general, to determine the linearized perturbations in the region $\Sigma$ (or more generally in the Rindler wedge $R_B$), we only need to know the boundary behavior in the domain of dependence region $D_B$, as discussed in detail in \cite{Almheiri:2014lwa}. The relevant boundary behaviour of each bulk field is captured by the one-point function of the corresponding operator. We can express the results as
\be
\label{bulktoboundary}
(\delta \phi_1)_\alpha(x,z)|_\Sigma = \int_{D_B} d^d x' K_\alpha(x,z ; x') \langle {\cal O}_\alpha(x') \rangle_{CFT}
\ee
where $K_\alpha(x,z ; x')$ is the relevant bulk-to-boundary propagator. As discussed in \cite{Morrison:2014jha, Almheiri:2014lwa,Bousso:2012mh}, $K_\alpha$ should generally be understood as a distribution to be integrated against consistent CFT one-point functions, rather than a function. Since the expression (\ref{bulktoboundary}) is linear in the CFT expectation values, the result (\ref{secondorder2}) is quadratic in these one-point functions and represents our desired second-order result.

To summarize, for a holographic state, the second-order contribution to entanglement entropy in the expansion (\ref{Sexp}) is the leading order contribution to the relative entropy $S(\rho_B||\rho_B^{vac})$. This is dual to canonical energy, given explicitly by:
\be
\label{Stwo}
\delta^{(2)} S_B = -\langle \delta \rho_1, \delta \rho_1 \rangle_{\rho^{vac}_B} = - \half {\cal E}( \delta \phi_1, \delta \phi_1) = -\half \int_\Sigma \omega(\delta g_1, \lie{\xi_B} \delta g_1) + \frac{1}{2} \int_\Sigma \xi^a_B T^{(2)}_{ab} \epsilon^b \; .
\ee
This is quadratic in the linearized perturbations $\delta \phi_\alpha$ (including the metric perturbation, and these can be expressed in terms of the CFT one-point functions on $D_B$ as (\ref{bulktoboundary}).

\subsection{Example: CFT$_2$ stress tensor contribution}
\label{SubSec:CFT2stress}

In this section, as a sample application of the general formula, we provide an explicit calculation of the quadratic stress tensor contribution to the entanglement entropy for holographic states in two-dimensional conformal field theories. This arises from the first term in (\ref{secondorder2}).

For a general CFT state, the stress tensor is traceless and conserved,
\be
\langle T^\mu {}_\mu \rangle = \langle \partial_\mu T^{\mu \nu} \rangle = 0 \; .
\ee
In two dimensions, these constraints can be expressed most simply using light-cone coordinates $x^\pm = x \pm t$, where we have
\be
\langle T_{+-} \rangle = \partial_+ \langle T_{--} \rangle = \partial_- \langle T_{++} \rangle = 0 \; .
\ee
Thus, a general CFT stress tensor can be described by the two functions, $\langle T_{++} (x^+)\rangle$ and $\langle T_{--} (x^-)\rangle$.

Assuming that the state is holographic, there will be some dual geometry of the form (\ref{FG}). According to (\ref{eq:HoloRenT}), the stress tensor expectation values determine the asymptotic form of the metric as
\be
\Gamma_{++}(x,0) = 8 \pi {G_N \over \ell_{AdS}} \langle T_{++} (x^+)\rangle \qquad \Gamma_{--}(x,0) = 8 \pi {G_N \over \ell_{AdS}} \langle T_{--} (x^-)\rangle
\ee
Now, suppose that our state represents a small perturbation to the CFT vacuum, so that the stress tensor expectation values and the asymptotic metric perturbations are governed by a small parameter $\lambda$:
\be
\label{boundary}
\Gamma_{++}(x,0) \equiv \lambda h_+(x^+) \qquad \Gamma_{--}(x,0) \equiv \lambda h_-(x^-) \; .
\ee
Then the metric perturbation throughout the spacetime is determined by this asymptotic behavior by the Einstein equations linearized about AdS. Here, we need only the components in the field theory directions, which give
\be
{1 \over z^3} \partial_z (z^3 \partial_z \Gamma_{\mu \nu}) + \partial_\rho \partial^\rho \Gamma_{\mu \nu} = 0 \; .
\ee
The solution in our Fefferman-Graham coordinates with boundary behaviour (\ref{boundary}) is
\be
\Gamma^{(1)}_{++}(x,z) = \lambda h_+(x^+) \qquad \Gamma^{(1)}_{--}(x,z) = \lambda h_-(x^-) \;
\ee
with the linearized perturbation $\Gamma^{(1)}_{\mu \nu}$ independent of $z$.

\subsubsection*{Satisfying the gauge conditions}

We would now like to evaluate the metric contribution to (\ref{Stwo})
\be
\label{S2grav}
\delta^{(2)}S_B^{grav} = - {1 \over 2} \int_\Sigma \omega^{grav}(\delta g_1, \lie{\xi_B} \delta g_1) \; .
\ee
This formula assumes the gauge conditions (\ref{eq:gauge-condition}) which differ from the Fefferman-Graham gauge conditions we have been using so far. Thus, we must find a gauge transformation to bring our metric perturbation to the appropriate form. In general, we can write
\begin{align}
\gamma_{ab}=h_{ab} +(\lie{V}g)_{ab} = h_{ab} + \nabla_a V_b + \nabla_b V_a \; .
\end{align}
where $\gamma$ is the desired metric perturbation satisfying the gauge condition, and $h$ is the perturbation in Fefferman-Graham coordinates (equivalent to $\Gamma$ for $d=2$).

The procedure for finding an appropriate $V$ and evaluating (\ref{S2grav}) is described in detail in \cite{Lashkari:2015hha}, but we review the main points here. Defining coordinates $(X^A, X^i)$ so that the extremal surface lies at some fixed value of $X^A$ with $X^i$ describing coordinates along the surface, the gauge condition (\ref{eq:gauge-condition}) (equivalent to requiring that the coordinate location of the extremal surface remains fixed) gives
\be
\label{eq:V-eqs0}
(\nabla_i \nabla^i V_A + [\nabla_i, \nabla_A] V^i + \nabla_i h^i_A - {1 \over 2} \nabla_A h^i {}_i)|_{\tilde{B}} = 0
\ee
while the condition (\ref{eq:gauge-condition2}) that $\xi_B$ continues to satisfy the Killing equation at $\tilde{B}$ gives
\begin{align}
	\left( h_{iA} + \nabla_i V_A + \nabla_A V_i \right)   |_{\tilde B } &= 0 \, , \label{eq:V-eqs1}\\
	\left( h^A_{\,\,D} - \frac{1}{2} \delta^A_{\,\,D} h^C_{\,\,C} + \nabla^A V_D + \nabla_D V^A - \delta^D_{\,\, D} \nabla_C V^C_{\,\,C} \right)\bigg|_{\tilde B} &= 0 \, .\label{eq:V-eqs2}
\end{align}

To solve these, we first expand our general metric perturbation in a Fourier basis.
\be
\label{fourier}
  h_{\mu\nu}(t,x, z)  =\lambda\int \left[  \delta_\mu^+ \delta_{\nu}^+ \hat h_+(k)  e^{i k x^+} + \delta_\mu^- \delta_{\nu}^-  \hat h_-(k) e^{i k x^-}  \right] d k\,,
\ee
with a gauge choice $h_{za } (t,x,z)= 0$.

For each of the basis elements, we use the equations (\ref{eq:V-eqs0}), (\ref{eq:V-eqs1}) and (\ref{eq:V-eqs2}) to determine $V$ and its first derivatives at the surface $V$. For these calculations, it is useful to define polar coordinates $(z,x) = (r \cos \theta, r \sin \theta)$.
Since the gauge conditions are linear in $V$, the conditions on $V$ for a general perturbation are obtained from these by taking linear combinations as in (\ref{fourier}),
\begin{align}
  V_{a} (t,x,z) & =\lambda\int  \left[ \hat{V}_{a}^{+}(k) e^{i k x^+} + \hat{V}_{a}^{-}(k)    e^{i k x^-} \right] dk \,\,.
\end{align}

After requiring $V_a$ remain finite at $\theta = \pm \frac{\pi}{2}$, we find
\begin{small}
\begin{align}
\hat{V}_t^-(k;t,r,\theta) &=\frac{e^{-i k t}}{k^3 r^2 \cos^2\theta} \left(-i\cos(kr) + \sin\theta\sin(kr)
-i\frac{(k^2r^2\cos^2\theta-1)e^{i k r\sin\theta}}{2}\right) \nonumber \\
\hat{V}_r^-(k;t,r,\theta) &=\frac{e^{-i k t}}{k^3r^2\cos^2\theta}\Bigg( \sin(kr) - i\sin\theta\cos(kr) \nonumber \\
& \qquad\qquad\qquad-\frac{(k^2r^2\cos^2\theta\sin\theta+ikr\cos^2\theta+2i\sin\theta)e^{i k r \sin\theta}}{2}\Bigg) \nonumber\\
\partial_t\hat  V^-_{\theta} (k;t,r,\theta)& =
\frac{e^{-ikt}}{2\,k^{2}\,r\cos\theta}
\left(
		(2 + k^{2}r^{2}\cos^{2}\theta-2\,i k r\,\sin\theta)e^{ikr\sin\theta} - \frac{2\sin(kr)}{k^3r^2}
\right)\,\nonumber
\\
\partial_r \hat V^-_{\theta} (k;t,r,\theta)& =
\frac{e^{-ikt}}{k^{3}r^{2}\cos\theta}\Big(
2i\cos(kr) \nonumber\\
&
\quad+ \left[2kr\sin\theta+r^{3}k^{3}\sin\theta\cos^{2}\theta+i\left(r^{2}k^{2}\cos^{2}\theta-kr^{2}+2\right)\right]e^{ikr\sin\theta}  \Big)\nn
\end{align}
where the $V^{\pm}$ solutions are related through $\hat V_{r}^{+}(k;t, r,\theta)=\hat V_{r}^{-}(k;-t,r,\theta)$ and $\hat V_{t}^{-}(k;t,r,\theta)=-\hat V_{t}^{+}(k;-t,r,\theta)$.
\end{small}
The results here give the behavior of $V$ and its derivatives only at the surface $\tilde{B}$ ($r=R$ in polar coordinates). Elsewhere, $V$ can be chosen arbitrarily, but we will see that our calculation only requires the behavior at $\tilde{B}$.

\subsubsection*{Evaluating the canonical energy}

Given the appropriate $V$, we can evaluate (\ref{S2grav}) using
\bea
\label{omega}
\omega(g,\gamma,\lie{\xi} \gamma) &=& \omega(h + \lie{V} g, \lie{\xi_B} (h + \lie{V} g)) \\
&=& \omega(g,h,\lie{\xi} h) + \omega(g,h + \lie{V} g,\lie{[\xi,V]} g) - \omega(g,\lie{\xi} h,  \lie{V} g) \nonumber
\eea
where
\be
[\xi,V]^a = \xi^b \partial_b V^a - V^b \partial_b \xi^a \;
\ee
and we have used that $\lie{\xi} g = 0$. We can simplify this expression using the gravitational identity
\be
\label{dchi}
\omega(g, \gamma,\lie{\xi} g) = d \chi(\gamma,X)
\ee
where
\be
\label{defchi}
\chi(\gamma,X) = {1 \over 16 \pi G_N} \epsilon_{ab} \left\{\gamma^{ac} \nabla_c X^b - {1 \over 2} \gamma_c{}^c \nabla^a X^b + \nabla^b \gamma^a {}_c X^c - \nabla_c \gamma^{ac} X^b + \nabla^a \gamma^c {}_c X^b \right\} \; .
\ee
Thus, we have
\be
\label{defomega}
\omega(g,\gamma,\lie{\xi} \gamma) = \omega(g,h,\lie{\xi} h) + d \rho
\ee
where
\be
\label{defrho}
\rho = \chi(h + \lie{V} g,[\xi,V])  - \chi(\lie{\xi} h,V)  \; .
\ee
Finally, choosing $V$ so that it vanishes at $B$, we can rewrite (\ref{S2grav}) as
\be
\label{calcE}
{\cal E} = \int_\Sigma \omega(g, h, \lie{\xi} h) + \int_{\tilde{B}} \rho(h,V) \; .
\ee
In this final expression, we only need $V$ and its derivatives at the surface $\tilde{B}$.
Thus, we can now calculate the result explicitly for a general perturbation. In the Fourier basis, the final result in terms of the boundary stress tensor is
\begin{align}
 \label{eq:k-canonical-energy}
\mathcal {E}=& \int dk_1 \! \int dk_2 \,
\hat K_2(k_1, k_2)\,
 \langle T_{++}(k_1) \rangle \langle T_{++} (k_2) \rangle + \{+ \leftrightarrow - \}\,,
\end{align}
where the kernel is
\begin{small}
\begin{align}
\label{eq:canon-kernel}
\hat K_2(k_1, k_2) = \frac{256 \pi^2\, R^4 \,G_N}{\ell_{AdS} K^{3}(K-\kappa)^{3}(K+\kappa)^{3}}
&\left[(K^{5}-2\,(\kappa^{2}+4)K^{3} +\kappa^{4}K)\cos K
\right. \nn & \left.
-(5K^{4}-6K^{2}\kappa^{2}+\kappa^{4})\sin K
+8\,K^{3}\cos\kappa \right] \;,
\end{align}
\end{small}
$\!\!$with $K\equiv R(k_1+k_2), \kappa \equiv R(k_1-k_2)$. We note in particular that the result splits into a left-moving part and a right-moving part with no cross term.

Transforming back to position space
\begin{align}
 \mathcal{E}  &=   \int_{B'} dx_1^+  \! \int_{B'} dx_2^+ \,  K_2(x_1^+, x_2^+)   \, \la T_{++}(x_1^+)  \ra \la T_{++} (x_2^+) \ra   + \{+ \leftrightarrow - \}\,,
\end{align}
where the kernel $K_2$ is symmetric under exchange of $x_1^\pm$ and $x_2^\pm$, and has support only on $x_i^\pm \in[-R,R]$. Focusing only on the domain of support, we have
\begin{align}
K_2(x_1, x_2) &= \frac{4 \pi^2 G_N }{R^{2} \ell_{AdS}} \begin{cases}
 (R-x_1)^2 (R+x_2)^2& x_1\geq x_2 \\
 (R+x_1)^2 (R-x_2)^2 & x_1 < x_2
\end{cases} \,.
\label{eq:CanEnergyKernel}
\end{align}
Using the relation $c = 3 \ell_{AdS} / (2 G_N)$ between the CFT central charge and the gravity parameters, we recover the result (\ref{result2D}) from the introduction.

Like the leading order result in (\ref{Sexp}), the integrals can be taken over any surface $B'$ with boundary $\partial B$. The fact that we only need the stress tensor on a Cauchy surface for $D_B$ is special to the stress tensor in two dimensions, since the conservation relations allow us to find the stress tensor expectation value everywhere in $D_B$ from its value on a Cauchy surface. For other operators, or in higher dimensions, the result will involve integrals over the full domain of dependence. We will see an explicit example in the next subsection.

Positivity of relative entropy requires $\E$ to be positive which requires the kernel to be positive semi-definite. As we show in Appendix \ref{appendix:positivity}, one can demonstrate that the positivity explicitly, providing a check of our results. An alternative proof of positivity is given in Section \ref{Sec:EntanglementHolography}. As a more complete check, we will show in Section \ref{Sec:CFT2} that this result can be reproduced by a direct CFT calculation for the special class of states that can be obtained from the vacuum state by a local conformal transformation.

\subsection{Example: Scalar operator contribution}
\label{sec:scalar}
We now consider an explicit example making use of the bulk matter field term in (\ref{secondorder2}) in order to calculate the terms in the entanglement entropy formula quadratic in the scalar operator expectation values. The discussion for other matter fields would be entirely parallel. This example is more representative, since the formula will involve scalar field expectation values in the entire domain of dependence $D_B$, i.e. a boundary spacetime region rather than just a spatial slice. The results here are similar to the recent work in \cite{Faulkner:2014jva, Speranza:2016jwt, Faulkner:2015csl}, but we present them here to show that they follow directly from the canonical energy formula.

We suppose that the CFT has a scalar operator of dimension $\Delta$ with expectation value $\langle {\cal O}(x) \rangle$. According to the usual AdS/CFT dictionary, this corresponds to a bulk scalar field with mass $m^2 = \Delta(\Delta - d)$ and asymptotic behavior
\be
\label{scalarboundary}
\phi(x,z) \to \gamma z^\Delta \langle {\cal O}(x) \rangle \; ,
\ee
where $\gamma$ is a constant depending on the normalization of the operator ${\cal O}$. The leading effects of the bulk scalar field on the entanglement entropy (\ref{Stwo}) come from the matter term in the canonical energy
\be
\label{StwoM}
\delta^{(2)} S^{matter}_B = \frac{1}{2}   \int_\Sigma \xi^a_B T^{(2)}_{ab} \epsilon^b \; .
\ee

Using the explicit form of $\xi_B$ from (\ref{defxi}) and $\epsilon$ from (\ref{defeps}), this gives (for a ball centered at the origin)
\be
\label{StwoM1}
\delta^{(2)} S^{matter}_B=  - \frac{\ell_{AdS}^{d-1}}{2} \int_0^R {dz \over z^{d-1}} \int_{x^2 < R^2 - z^2} d^{d-1} x {\pi \over R} (R^2 - z^2 - x^2)  T^{(2)}_{00}(x,z) \; .
\ee
This expression is valid for a general bulk matter field. For a scalar field, we have
\be
T^{(2)}_{ab} = \partial_a \phi_1 \partial_b \phi_1 - {1 \over 2} g_{ab} (g^{cd} \partial_c \phi_1 \partial_d \phi_1 + m^2 \phi_1^2) \;,
\ee
where $g_{ab}$ is the background AdS metric and $\phi_1$ represents the solution of the linearized scalar field equation on AdS,
\be
\label{scalareq}
{1 \over z^{d-1}} \partial_z \left\{ z^{d-1} \partial_z \phi \right\} + \partial_\mu \partial^\mu \phi - {m^2 \over z^2} \phi = 0\;,
\ee
with boundary behavior as in (\ref{scalarboundary}). This solution is given most simply in Fourier space, where we have
\be
\label{BBfour}
\phi_1(k,z) =  \frac{2^\nu \Gamma(\nu +1)}{(2\pi)^d }\int_{k_0^2 > \vec{k}^2  } d^d k \frac{e^{i k_\mu x^\mu}}{\left(k_0^2 - \vec k^2\right)^{\nu/2}}  \, z^{d \over 2} J_{\nu}\left(\sqrt{k_0^2 - \vec{k}^2} z \right) \, \gamma \langle {\cal O}(k) \rangle \; ,
\ee
where $\nu = \Delta - d/2$, but we can formally write a position-space expression using a bulk-to-boundary propagator $K(x, z ; x')$ as~\cite{Hamilton:2006az, Hamilton:2006fh}
\be
\label{BBpos}
\phi_1(x,z) = \gamma \int dx' K(x, z ; x') \langle {\cal O}(x') \rangle \; .
\ee
The integral here is over the boundary spacetime, however it has been argued (see, for example \cite{Almheiri:2014lwa,Morrison:2014jha}) that to reconstruct the bulk field throughout the Rindler wedge $R_B$ (and specifically on $\Sigma$), we need only the boundary values on the domain of dependence region. We recall some explicit formulae for this ``Rindler bulk reconstruction'' in Appendix \ref{app:Rindler_reconstruct}. Combining these results, we have a general expression for the scalar field contribution to entanglement entropy at second-order in the scalar one-point functions,
\bea
\label{StwoM2}
\delta^{(2)} S^{scalar}_B &=&  -\frac{\ell_{AdS}^{d-1}}{2} \int_0^R {dz \over z^{d-1}} \int_{x^2 < R^2 - z^2}  d^{d-1} x {\pi \over R} (R^2 - z^2 - x^2)   \\
&&  \qquad \qquad \qquad \qquad \left\{(\partial_0 \phi_1)^2 + (\partial_i \phi_1)^2 + (\partial_z \phi_1)^2 + {m^2 \over z^2} \phi_1^2 \right\} \nonumber
\eea
where $\phi_1$ is given in (\ref{BBfour}) or (\ref{BBpos}) .

As a simple example, consider the case where the scalar field expectation value is constant. In this case it is simple to solve (\ref{scalareq}) everywhere to find that
\be
\phi_1(x,z) = \gamma \langle {\cal O} \rangle z^\Delta \; .
\ee
Inserting this into the general formula (\ref{StwoM2}), and performing the integrals, we obtain
\be
\delta^{(2)} S^{scalar}_B = -{\pi  \ell_{AdS}^{d-1} \over 4 } \gamma^2 \langle {\cal O} \rangle^2 R^{2 \Delta} \Omega_{d-2} {\Delta \Gamma({d \over 2} - {1 \over 2}) \Gamma( \Delta - {d \over 2} + 1) \over \Gamma(\Delta + {3 \over 2})}\,.
\ee
This reproduces previous results in the literature~\cite{Blanco:2013joa, Speranza:2016jwt}.


\section{Stress tensor contribution: direct calculation for CFT$_2$}
\label{Sec:CFT2}

In Section \ref{SubSec:CFT2stress}, we used the equivalence between quantum Fisher information and canonical energy to obtain an explicit expression for the second-order stress tensor contribution to the entanglement entropy for holographic states in two-dimensional CFTs. This is applicable for general holographic states, whether or not other matter fields are present in the dual spacetime (in which case there are additional terms in the expression for entanglement entropy). In special cases where there are no matter fields, the spacetime is locally AdS and we can understand the dual CFT state as being related to the vacuum state by a local conformal transformation. We show in this section that in this special case, we can reproduce the holographic result (\ref{eq:CanEnergyKernel}) through a direct CFT calculation, providing a strong consistency check. We note that the result does not rely on taking the large $N$ limit or on any special properties of the CFT, so the formula holds universally for this simple class of states.

Our approach will be to develop an iterative procedure to express the entanglement entropy as an expansion in the stress tensor expectation value for this special class of states.
 We evaluate the entanglement entropy for these states from a correlation function of twist operators obtained by transforming the result for the vacuum state.\footnote{A similar approach was recently used to derive the modular Hamiltonian of these excited states in~\cite{Lashkari:2015dia}.} Similarly, the stress tensor expectation values follow directly from the form of the conformal transformation. Inverting the relationship between the required conformal transformation and the stress tensor expectation value allows us to express the entanglement entropy as a perturbative expansion in the expectation value of the stress tensor. Similar CFT calculations have also been used recently in \cite{helleretal}.

\subsection{Conformal transformations of the vacuum state}
In two-dimensional CFT, under a conformal transformation $w = f(z)$, the stress tensor transforms as
\bea
T'(w) =   \left(\frac{d w}{d z}\right)^{-2} \left( T (z) + \frac{c}{12} \{f(z); z\} \right)\,.
\eea
Here $c$ is the central charge of the CFT and the inhomogeneous part is the Schwarzian derivative
\be
\{f(z); z\} \equiv \frac{f'''(z)}{f'(z)} -\frac{3 f''(z)^2}{2f'(z)^2}\,.
\ee
For an infinitesimal transformation $f(z) = z +  \lambda\, \epsilon(z)$,
the Schwarzian derivative can be expanded as
\begin{small}
\be
\{z+ \lambda  \epsilon(z) ; z \} = \lambda\,  \epsilon'''(z) -\lambda^2 \left( \epsilon'''(z) \epsilon'(z) + \frac{3}{2} \epsilon''(z)^2 \right) + \lambda^3\left(  \epsilon'(z)^2 \epsilon'''(z)+ 3\epsilon'(z) \epsilon''(z)^2 \right) + \cdots
\label{eq:PerturbativeSchwarzian}
\ee
\end{small}
$\!\!$The CFT vacuum is invariant under the $SL(2 ,\mathbb{C})$ subgroup of global conformal transformations.
However, for transformations which are not part of this subgroup, the vacuum state transforms into \emph{excited states}.  The action of the full conformal group includes the full Virasoro algebra which involves arbitrary products and derivatives of the stress tensor
\be
\text{Id} \sim 1, T, \partial^m T, T^2, T\partial^n T, \cdots \,.
\ee
These states capture the gravitational sector of the gravity dual. Other excited states can be obtained by the action of other primary operators and their descendants.
However we restrict ourselves to the class states that are related to `pure gravity' excitations, which are the states obtained by conformal transformation of the vacuum state.

We denote the excited state as
$| f \rangle = U_f \,| 0 \rangle$
where $U_f$ is the action of a conformal transformation on the vacuum $|0 \rangle$.
The expectation value of the stress tensor for the state perturbed state $|f\rangle$ is
\bea
\langle f | T(z) | f \rangle = \langle 0 | U_f^\dagger \, T(z)  \,U_f | 0\rangle =\langle 0 | T'(w) | 0\rangle = \left(\frac{df}{dz}\right)^{-2} \frac{c}{12} \{f(z) ; z\} \,,
\label{expT}
\eea
where we used that $ \langle 0| T(z)|0 \rangle =0$. The anti-holomorphic component of the stress tensor $\bar T(\bar z)$ is similarly related to the anti-holomophic part of the conformal transformation $\bar f$.

To leading order in a conformal transformation near the identity, this equation relates the conformal transformation to $\langle T(z) \rangle$ by a third-order ordinary differential equation. The three integration constants correspond to the invariance of $\langle T(z) \rangle$ under the global conformal transformations. Thus we have an invertible relationship between the conformal transformations modulo their global part and $\langle T(z) \rangle$, at least near the identity.

\subsection{Entanglement entropy of excited states }
In a two-dimensional CFT, the entanglement entropy can be explicitly computed using the replica method~\cite{Holzhey:1994we, Calabrese:2004eu}.
The computation can be reduced to a correlation function of twist operators $\Phi_\pm$, which are conformal primaries with weight $(h_n, \bar h_n) = \frac{c}{24}(n -1/n, n -1/n)$.

The R\'enyi entropy is
\be
\exp\left((1-n)S^{(n)} \right) = \langle  \Phi_+ (z_1) \Phi_- (z_2) \rangle = (z_2 - z_1)^{-2 h_n}\,.
\ee
The entanglement entropy is obtained by taking the $n\rightarrow 1 $ limit of $S^{(n)}$.
\be
S_{\text{vac}}   = \lim_{n \rightarrow 1} S^{(n)} = \lim_{n \rightarrow 1}  \, (1-n)^{-1} \log (z_2 -z_1)^{-2h_n} = \frac{c}{12} \log \frac{(z_2 - z_1)^2}{\delta^2} \,.
\ee
For the excited states obtained by conformal transformations $z \rightarrow w = f(z)$\, the R\'enyi entropy is
\bea
\exp\left((1-n)S_{\text{ex}}^{(n)} \right) &=& \langle f | \Phi_+ (z_1) \Phi_- (z_2) | f \rangle\\
&=&
\left(\frac{df}{dz}\right)^{-h_n}_{z_1} \left(\frac{df}{dz}\right)^{-h_n}_{z_2}
\left(\frac{d\bar f}{d\bar z}\right)^{-\bar h_n}_{\bar z_1}
\left(\frac{d\bar f}{d\bar z}\right)^{-\bar h_n}_{\bar z_2}
\langle 0 | \Phi_+ (z_1) \Phi_- (z_2) | 0 \rangle\,. \nn
\eea
Here $z_1, z_2$ are the points $f(z_1) = \bar f (\bar z_1) = -R$, $f(z_2) = \bar f(\bar z_2) = R$.
The entanglement entropy of the excited state is
\be
S_{\text{ex}}  = \lim_{n \rightarrow 1} S_{\text{ex}}^{(n)}
= \frac{c }{12} \log \left| \frac{
f'(z_1) f'(z_2)  \bar f'(\bar z_1) \bar f'(\bar z_2)
(z_2 -z_1)^2 }{\delta^2} \right| \,.
\ee
Therefore the change in entanglement entropy respect to the vacuum state is
\begin{align}
\label{deltaSEE}
\delta S \equiv S_{\text{ex}} - S_{\text{vac}}
	=& \frac{c}{12} \log \left| \frac{f'(f^{-1}(R)) f'(f^{-1}(-R))  (f^{-1}(R) -f^{-1}(-R))^2 }{(2R)^2} \right| \\
	&+  \frac{c}{12} \log \left| \frac{\bar f'(\bar f^{-1}(R)) \bar f'(\bar f^{-1}(-R))  (\bar f^{-1}(R) -\bar f^{-1}(-R))^2 }{(2R)^2} \right|
	 \,. \nonumber
\end{align}

By inverting   \eqref{expT}, the conformal transformation required to reach the state $| f \rangle$ can be expressed as a function of the expectation value of the stress tensor. Plugging this $f$ into \eqref{deltaSEE}, allows us to express the entanglement entropy as a function of the expectation value of the stress tensor alone, as we set out to do.

In practice, we will invert   \eqref{expT} order by order in a small conformal transformation and express the entanglement entropy as an expansion in the resulting small stress tensor. The second-order term in this expansion will be the Fisher information metric.

In the following, we will focus on the holomorphic term in \eqref{expT}, noting that the anti-holomorphic part follows identically.\footnote{Note that the potential cross-term between left and right moving contributions vanished in the gravitational computation of $\delta^{(2)} S$.}

\subsection{Perturbative expansion}

Consider a conformal transformation perturbation near the identity transformation
\be
w = f(z) = z + \lambda f_1 (z) + \lambda^2 f_2(z) + \lambda^3 f_3(z) + \cdots\,,
\ee
where $\lambda$ is a small expansion parameter.

In this expansion,
\begin{small}
\be
\frac{12}{c} \, \langle T(w) \rangle =  \lambda\, f_1'''(w)  + \lambda^2 \left( -\frac{3}{2}f_1''(w)^2 -3f_1'(w)f_1'''(w)+f_2'''(w)-f_1(w)f_1''''(w)\right) + {\cal O}(\lambda^3)\,,
\label{stressPerturbation}
\ee
\end{small}
$\!\!$and the entanglement entropy is
\begin{small}
\begin{align}
\frac{12}{c} \, S_{\text{ex}}  =&  \log \left| \frac{f'(z_1) f'(z_2)(z_2 - z_1)^2}{\delta^2} \right |\nn
				=& \log \frac{ (2R)^2}{\delta^2}  +\lambda\left[\frac{R \left(f_1'(-R)+f_1'(R)\right)+f_1(-R)-f_1(R)}{R}\right]\nn
				&+ \lambda^2 \Big(-\frac{(f_1(R)-f_1(-R))^2}{4R^2} + \frac{-f_1(-R) f_1'(-R)+f_1(R) f_1'(R)+f_2(-R)-f_2(R)}{R}  \nn
				& \phantom{+ \lambda^2 ( ( }-\frac{1}{2} f_1'(-R){}^2-\frac{1}{2} f_1'(R){}^2+f_2'(-R)+f_2'(R)-f_1(-R) f_1''(-R)-f_1(R) f_1''(R) \Big) \nn
				&+ \,{\cal O}(\lambda^3) \,. \label{perturbedSEE}
\end{align}
\end{small}
\subsubsection*{Linear order}
To first-order in $\lambda$, the stress tensor is given by
\be
\langle T(z) \rangle = \lambda \frac{c}{12}  \,  f_1 '''(z) + {\cal O}(\lambda^2) \,,
\ee
so that change in the expectation value of  the modular Hamiltonian becomes
\bea
\delta \langle H_B \rangle &=&  \frac{\lambda\, c}{24 R}    \int_{-R}^{R} dz \,  (R^2 - z^2) f_1'''(z) \nn
					&=& \frac{\lambda\, c }{24R} \left[ (R^2-z^2) f_1''(z) + 2\left( z f_1'(z) - f_1(z) \right) \right]^{R}_{-R}   \nn
					&=& \frac{\lambda\, c }{12 R}  \left[R( f_1'(R) + f_1'(-R)) - ( f_1(R) - f_1(-R) )\right] \,.
\label{eq:ModHChange}
\eea
From \eqref{deltaSEE} we also have that the first-order change in entanglement entropy is
\bea
\delta^{(1)} S
	= \frac{\lambda\, c }{12 R} \left[R( f_1'(R) + f_1'(-R)) - ( f_1(R) - f_1(-R) )\right] \,.
\eea
Comparing with \eqref{eq:ModHChange} we see that the first law of entanglement holds
\be
\delta^{(1)} S = \delta \langle H_B \rangle   \,.
\ee

\subsubsection*{Second-order}
The second-order change in entanglement entropy gives the second-order relative entropy as the modular Hamiltonian is linear in the expectation value of the stress tensor. This is the quantum Fisher metric in the state space, which is dual to the canonical energy in gravity~\cite{Lashkari:2015hha}. In this section, we obtain the expression for canonical energy from the CFT side and find an exact match to the results of Section \ref{SubSec:CFT2stress}.

Our procedure so far yields the entanglement entropy of a subregion in terms of a perturbative expansion in small stress tensor expectation value
\bea
\delta S = && \int_B \frac{d z}{2\pi} \, K_1 (z) \langle  T(z) \rangle - \frac{1}{2} \int_B  \frac{d z_1}{2\pi} \, \int_B \frac{d z_2}{2\pi} \, K_2 (z_1, z_2) \langle T(z_1) \rangle\langle T(z_2) \rangle +\cdots \nn
&&+ \, \{ z \leftrightarrow \bar z\} \,.
\label{eq:SEEKernelCFT}
\eea

To obtain $K_2(z_1,z_2)$, we need to invert the relationship in \eqref{stressPerturbation} order by order, the lower order solutions $f_{i-1}, f_{i-2}, \cdots f_1$ becoming sources for the $i$-th order solution.

Taking the explicit expression for $\langle T (z) \rangle$ to simplify solving the differential equations,
\be
\langle T (z) \rangle = \lambda \, \left( c_1 e^{i k_1 z } +   c_2 e^{i k_2 z } \right)\,,
\ee
is sufficient to extract the Fourier transformed kernel.

The first-order solution is
\be
f_1(z) =  F_1 + F_2 z + F_3 z^2 +  \frac{12 i}{c}\left( \,c_1 \frac{e^{i k_1 z}}{k_1^3}+  \,c_2 \frac{e^{i k_2 z}}{k_2^3}\right)\,,
\ee
where $F_i$ are constants that corresponds to the global part of the conformal transformation and do not effect the final result. We take these constants to be zero for simplicity.
The second-order solution is
\be
f_2(z) =  -\frac{9}{c^2}\left[ \frac{ 11 i}{16} (c_1^2 \frac{e^{2 i  k_1 z } }{k_1^5} + c_ 2^2 \frac{e^{2 i k_2 z}}{k_2^5}  )  + i \frac{ c_1 c_2  }{ k_1^3 k_2^3}\frac{ e^{i(k_1+k_2) z}   \left(k_1^4+3 k_2 k_1^3+3 k_2^2 k_1^2+3 k_2^3 k_1+k_2^4\right)}{(k_1 +k_2)^3} \right] \,.
\ee

With these solutions, we obtain
\bea
\tilde K_1 ( k) =&&   \frac{2}{k^2} \, \frac{\sin \left(k R\right)-k R \cos \left(k R\right)}{k  R}\,,
\eea
as well as
\begin{small}
\begin{align}
\label{eq:CFTk-kernel}
 \tilde K_2(k_1, k_2) &=\frac{96 R^4}{c}  \frac{ (K^{5}-2(\kappa^{2}+4)K^{3}+\kappa^{4}K)\cos K
-(5K^{4}-6K^{2}\kappa^{2}+\kappa^{4})\sin K+8K^{3}\cos\kappa}{ K^{3}(K-\kappa)^{3}(K+\kappa)^{3}} \,,
\end{align}
\end{small}
$\!\!$with $K \equiv R(k_1 + k_2)$ and $ \kappa \equiv R(k_1 - k_2) $.

Taking the inverse Fourier transformation of $\tilde K_1 ( k) $
\bea
K_1(z) &=& \int d k  \, \tilde K_1 ( k) e^{-i k z}
	=\pi \frac{ R^2-z^2 }{R} W(R, z)
\eea
where
\be
W(R, x) \equiv \frac{\left(\text{sgn}\left(R+x\right)+ \text{sgn}\left(R-x\right)\right)}{2}
\ee
is a window function with support $x\in [-R,R]$.

The second-order position space kernel  is
\begin{align}
K_2(z_1, z_2)
&=\frac{6 \pi ^2 }{c R^2}  \begin{cases}
 (R-z_1)^2 (R+z_2)^2& -R \leq z_2\leq z_1 \leq R\\
 (R+z_1)^2 (R-z_2)^2 & -R \leq z_1 < z_2 \leq R
\end{cases} \,.
\label{eq:secondKernel}
\end{align}
The anti-holomorphic part is the same with $z\rightarrow\bar z$, and the cross term vanishes. With the relation
\be
c = \frac{3\ell_{AdS}}{2 G_N}
\label{eq:BrownHenneauxC}
\ee
this reproduces the kernel for canonical energy in   \eqref{eq:CanEnergyKernel}.

This result holds for regions defined on any spatial slice of the CFT. If we choose the $t=0$ slice, $z=\bar z=x$ and our result becomes
\bea
\delta S^{(2)} _{EE}  &=& - \half \int_B d x_1 \, \int_B d x_2 \, K_2 (x_1,  x_2) \left[ \langle T_{++}(x_1)\rangle \langle T_{++}(x_2)\rangle
+ \langle T_{--}(x_1)\rangle \langle T_{--}(x_2)\rangle \right] \,.\nonumber
\eea
Changing variables using $x_1=x-r$, $x_2=x+r$, the kernel is simply
\begin{align}
K_2(x,r)=K_2(x,-r)=\frac{12 \pi ^2 }{c R^2} \left[ (R-|r|)^2 -x^2 \right]^2 \Theta\left(R-|r|-|x|\right).
\end{align}

\subsection{Excited states around thermal background}
\label{sec:thermal}
A similar analysis can be applied to perturbations around a thermal state with temperature $ T = \beta^{-1}$. If we denote homogeneous thermal state $| \beta \rangle$, the stress tensor one-point function is
\be
\langle \beta | T | \beta \rangle = \frac{\pi^2 c}{6 \beta^2} \,.
\ee
This can be obtained by a conformal transformation from the vacuum with
\be
f_\beta ( z ) = \frac{\beta}{2 \pi} \log(z)\,.
\label{eq:ConfTransfToThermal}
\ee

On top of this transformation, one could also apply an infinitesimal conformal transformation to obtain non-homogeneous perturbation around thermal state.

A similar computation as the previous section leads to the first-order kernel
\be
K_1^\beta (z) =  \frac{2 \beta}{\sinh(\tfrac{2\pi R}{\beta})} \sinh\left(\frac{\pi(R-z)}{\beta}\right) \sinh\left(\frac{\pi(R+z)}{\beta}\right) \,,
\ee
which is the modular hamiltonian of thermal state in 2d CFT.

Furthermore, the second-order kernel is
\begin{align}
K^\beta_2(z_1, z_2)=\frac{24 \beta^2 }{c \, \sinh^2(\tfrac{2\pi R}{\beta})}  \begin{cases}
\sinh^2\left(\frac{\pi(R-z_1)}{\beta}\right)  \sinh^2\left(\frac{\pi(R+z_2)}{\beta}\right) & -R\leq z_2 \leq z_1 \leq R\\
\sinh^2\left(\frac{\pi(R+z_1)}{\beta}\right)  \sinh^2\left(\frac{\pi(R-z_2)}{\beta}\right) & -R\leq z_1 < z_2 \leq R
\end{cases} \,.
\label{secondThermalKernel}
\end{align}

\subsection*{Consistency check : homogeneous BTZ perturbation}

As a check, consider the homogeneous perturbation example, where $\langle T \rangle=\langle \bar T\rangle  = \frac{\lambda}{8 G_N}$.\footnote{$\lambda = \frac{2\pi^2}{\beta}$ sets the temperature.}
In AdS$_3$ this is a perturbation towards the planar BTZ geometry
\be
ds^2 = \frac{1}{z^2} \left(dz^2 + (1+\lambda z^2/2)^2 dx^2 - (1-\lambda z^2/2)^2 dt^2 \right)
\ee
in Fefferman-Graham coordinates.
Holographic renormalization \eqref{eq:HoloRenT} tells us the stress tensor expectation value of the dual CFT is
\be
\langle T_{tt}\rangle= \frac{1}{2\pi} \left( \langle T \rangle + \langle \bar T \rangle \right) = \frac{\lambda}{8 \pi G_N}\,.
\ee
As the black hole corresponds to the thermal state in CFT, the dual state be obtained by the conformal transformation~\eqref{eq:ConfTransfToThermal}.

First, applying \eqref{deltaSEE} for this conformal transformation, the change in entanglement entropy with respect to the vacuum is
\be
\delta S =\lambda \frac{ R^2 }{6G} - \lambda^2 \frac{R^4 }{90 G} + \lambda^3 \frac{4 R^6 }{2835 G} + \mathcal{O}(\lambda^4)\,,
\ee
which matches the previous known results \cite{Blanco:2013joa, Lashkari:2015hha}.

The linear order equals $\delta \langle H_B \rangle$ as expected from the entanglement first law.

The second-order term gives the quantum Fisher information or the canonical energy
\begin{align}
{\cal E} =\frac{d^2}{d\lambda^2}(\Delta E - \Delta S) \Big |_{\lambda =0 } = \frac{R^4}{45G_N}\,.
\end{align}
Using the formula using the second-order kernel \eqref{eq:SEEKernelCFT} and \eqref{eq:secondKernel},
we obtain the same canonical energy
\bea
{\cal E}  &=& 2\frac{d^2}{d\lambda^2} \left[\half \int_B \frac{d z_1}{2 \pi} \, \int_B \frac{d z_2}{2\pi} \, K_2 (x_1,  x_2)  \langle T\rangle \langle T\rangle \right]_{\lambda =0 } = \frac{R^4}{45G_N}\,.
\eea


\section{Auxiliary de Sitter space interpretation}
\label{Sec:EntanglementHolography}

In \cite{deBoer:2015kda}, it was pointed out that the leading order perturbative expression (\ref{firstorder}) for entanglement entropy, expressed as a function of the center point $x$ and radius $R$ of the ball $B$, is a solution to the wave equation for a free scalar field on an auxiliary de Sitter space, with $\langle T_{00}(x) \rangle$ acting as a source.

It was conjectured that higher order contributions might be accounted for by local propagation in this auxiliary space with the addition of self-interactions for scalar field.
In this section, we show that for two-dimensional CFTs, the second-order result (\ref{result2D}) can indeed be reproduced by moving to a non-linear wave equation with a simple cubic interaction to this scalar field. A slight complication is that we actually require two-scalar fields; one sourced by the holomorphic stress tensor $T_{++}$, and the other sourced by the anti-holomorphic part $T_{--}$; the perturbation to the entanglement entropy is then the sum of these two scalars, $\delta S = \delta S_+ + \delta S_-$, reproducing both terms in (\ref{result2D}).
We will focus on $\delta S_+$ since $\delta S_-$ follows identically.

To reproduce the second-order results for entanglement entropy, we consider an auxiliary de Sitter space with metric
\be
ds^2_{dS} = \frac{L_{{dS}}^2}{R^2}\left(-dR^2 + d  x^2\right)\,.
\ee
and consider a scalar field $\delta S_+$ with mass $m^2 L_{{dS}}^2= -2 $ and action
\begin{align}
{\cal L } =  \half \nabla_a \left( \delta S_+\right) \nabla^a \left(\delta S_+ \right)
+ \half m^2 \left(\delta S_+\right)^2
+ \frac{4}{c L_{{dS}}^2} \left(\delta S_+\right)^3 \,.
\label{eq:dSScalarAction}
\end{align}
The equation of motion is
\be
\left(  \nabla_{{dS}}^2- m^2 \right) \delta S_{+} (R,  x)  = \frac{12}{c L_{{dS}}^2} \left( \delta S_{+} (R,  x) \right)^2\,.
\label{eq:dSwaveEqInteraction}
\ee
As shown in \cite{deBoer:2015kda}, the first-order perturbation (\ref{firstorder}) obeys the linearized wave equation
\be
\left( \nabla_{{dS}}^2- m^2 \right) \delta^{(1)} S_+ (R,  x) = 0\,.
\ee
We can immediately check that the second-order perturbation (\ref{result2D}) is consistent with the nonlinear equation by acting with the dS wave equation on the second-order kernel \eqref{eq:secondKernel}
\be
\left( \nabla_{{dS}}^2 - m^2 \right) K_2 (x_1 -x, x_2- x) = - \frac{24}{c L_{{dS}}^2} K_1 ( x_1 - x) K_1 ( x_2 - x) \,.
\ee
Integration against the CFT stress tensor then gives \eqref{eq:dSwaveEqInteraction}.

\begin{figure}[h!]
  \centering
  \includegraphics{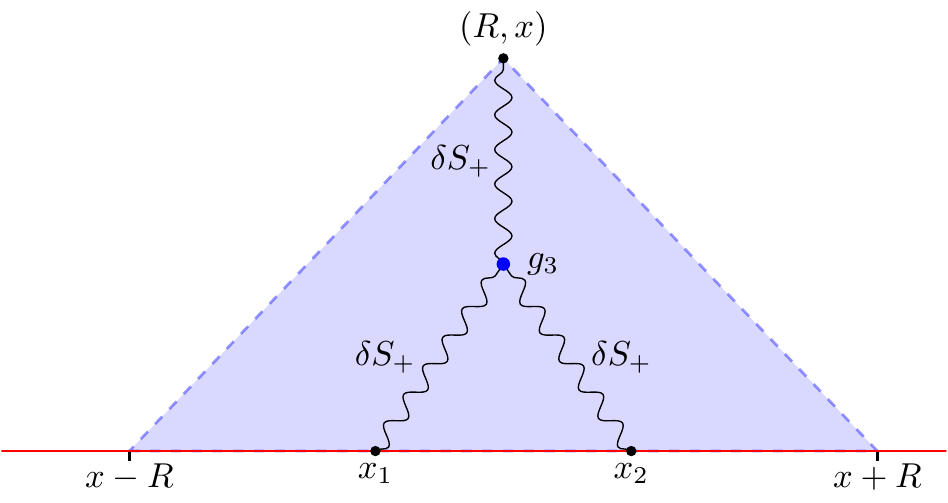}
  \caption{Feynman diagram which computes $\delta^{(2)} S$.  The $\delta S_+$ field propagates in de Sitter with a cubic interaction given by \eqref{eq:dSScalarAction}. The bold (red) line is the conformal boundary of de Sitter which is identified with a time slice of the CFT. $\delta S_+$ is sourced by the CFT stress tensor on this boundary. }
  \label{fig:ds-propagator}
\end{figure}

Alternatively, introducing the retarded\footnote{These propagators are defined to be non-zero only within the future directed light-cone. This is important in reproducing both the support and the exact form of $K_2(x_1,x_2)$. } bulk-to-bulk propagator~\cite{Xiao:2014uea}
\bea
G_{dS} (\eta , x ; \eta',  x \, ' ) = - \frac{\eta^2 + \eta\,'^2 -( x-  x \,')^2}{4 \eta \eta'}
\label{eq:dSbulktobulk}
\eea
and bulk-to-boundary propagator
\bea
K_{dS} (\eta ,  x;   x \, ' ) &=& \lim_{ \epsilon \rightarrow 0} \left [ -4 \pi \epsilon  \, \lim_{\eta'\rightarrow \epsilon}G_{dS} (\eta , x ; \eta',  x \, ' ) \right]  = \pi \frac{\eta^2 - \left( x -  x'\right)^2}{ \eta } \,,
\label{eq:dSboundarytobulk}
\eea
we can show directly that the solution with boundary behavior
\begin{align}
\delta S_+ = \frac{4 \pi}{3} \langle T_{++} \rangle R^2 + {\cal O}(R^3)\,.
\end{align}
For $ R \rightarrow 0$ gives
\begin{small}
\bea
\delta^{(1)} S_{+} (R, x_0) = \int dx \, K_{dS} (R, x_0; x) \langle T_{++} (x) \rangle
\eea
at first-order and
\bea
\delta^{(2)} S_{+} (R, x_0) =
\frac{12}{ c L_{dS}^2} \int_{dS} d\eta' dx' \sqrt{|g_{dS}|} \, G_{dS} (R,x_0; \eta' ,x')
\left(\int  dx \, K_{dS} (\eta' ,x';x) \langle T_{++} ( x) \rangle \right)^2 \, ,\nn
\label{EEfromdSTT}
\eea
\end{small}
$\!\!$at second-order, where the latter term comes from the diagram shown in Figure \ref{fig:ds-propagator}.

The integrals can be performed directly to show that these results match with the expressions (\ref{firstorder}) and (\ref{result2D}) respectively.

A useful advantage of writing the second-order result in the form (\ref{EEfromdSTT}) is that it is manifestly negative. More explicitly, we have
\bea
\delta^{(2)} S_+ (R, x_0) &=&
- \frac{3}{c L_{dS}^2} \int d\eta dy  \sqrt{|g_{{dS}}|} \, \frac{R^2 + \eta^2 -( x_0 -  y)^2}{R \, \eta}
\left[ \int_{B_y} dx \, K_{{{dS}}} (\eta , y ;x) \langle T_{++} ( x) \rangle \right]^2 \,.  \nn
\eea
where $\sqrt{|g_{{dS}}|}$ and the squared expression are manifestly positive and
the bulk-to-bulk propagator \eqref{eq:dSbulktobulk} is positive over the range of integration where $(y-x_0)^2 \leq (R-\eta)^2$. That this expression is negative is required by the positivity of relative entropy, since we showed above that $-\delta^{(2)} S$ represents the leading  order  perturbative expression for the relative entropy.

Recently, it has been realized that the modular Hamiltonian in certain non-vacuum states in two dimensional CFTs can be described by propagation in a dual geometry \cite{Asplund:2016koz} matching the kinematic space found previously in \cite{Czech:2014ppa,Czech:2015qta,Czech:2015xna,Czech:2015kbp}. We find that the results of Section \ref{sec:thermal} can be explained by the same interacting theory \eqref{eq:dSScalarAction} on this kinematic space. The kinematic space dual to the thermal state is
\footnote{The kinematic space dual to the BTZ black hole was first described in \cite{Czech:2014ppa,Czech:2015qta}. The explicit form of the metric in the coordinates we are using can be found in \cite{Asplund:2016koz}.}
\begin{align}
ds^2 = \frac{4 \pi^2 L_{dS}^2}{\beta^2 \sinh^2 \left( \frac{2 \pi R}{\beta} \right)} \left(-d R^2 + dx^2\right)\,.
\end{align}
The second-order perturbation to the entanglement entropy from \eqref{secondThermalKernel} obeys the wave equation \eqref{eq:dSwaveEqInteraction} 
 with the same interactions in this kinematic space.

We could imagine adding additional fields propagating in de Sitter to capture the contributions to the entanglement entropy from scalar operators discussed in Section \ref{sec:scalar}. However, unlike the contribution from the stress tensor, this contribution involves integration of the one-point functions over the full domain of dependence $D_B$. In higher-dimensions, this will also be true for the stress tensor contribution. The $R=0$ boundary of the auxiliary de Sitter space does not include the time direction of the CFT, so any extension of these results to contributions of other operators or higher dimensional cases will require a more sophisticated auxiliary space. Promising work in this direction is discussed in \cite{Czech:2016xec,helleretal}.


\section*{Acknowledgements}
We thank Michal Heller, Ali Izadi Rad, Nima Lashkari, Don Marolf, Robert Myers, and Philippe Sabella-Garnier for helpful discussions.
This research is supported in part by the Natural Sciences and Engineering Research Council of Canada, and by grant 376206 from the Simons Foundation.

\appendix


\section{Direct proof of the positivity}
\label{appendix:positivity}
Consider the left moving part of perturbation $h_+(x^+) \propto T_{++}(x^+)$.
The real space
$h_{+}(x)$ must be real valued functions for a perturbation of AdS$_{3}$.
We can expand $h_{+}(x)$ in a Taylor series $h_{+}(x)=\sum_{n=0}^{\infty}a_{n}x^{n}$
so that the canonical energy is given by
\begin{align}
\mathcal{E} \sim \sum_{n}\sum_{m}a_{n}a_{m}\int_B \int_B dx_{1}dx_{2}\,x_1^{n}x_2^{m}K_2(x_{1},x_{2}) \sim \sum_{n}\sum_{m}a_{n}a_{m}R^{n+m+4}\mathcal{A}_{n,m}  \,.\label{eq:left-int}
\end{align}
where the proportionality factor is up to a positive constant and
\begin{equation}
\mathcal{A}_{n,m}=\frac{1}{(n+m+3)(n+m+1)}\begin{cases}
0 & \text{if }n+m\text{ odd}\\
\frac{1}{(n+1)(m+1)} & \text{if \ensuremath{n,m} even}\\
\frac{nm+n+m+3}{nm(n+2)(m+2)} & \text{if \ensuremath{n,m} odd}
\end{cases}
\end{equation}
which is clearly non-negative and symmetric in $n,m$.

To show that the canonical energy is positive, we need to show the
matrix $M$ with entries given by $A_{n,m}=\mathcal{A}_{n-1,m-1}$
\footnote{The inelegant notation change is due to conventional matrix notation
starting at $n=1$, while the Taylor series starts at $n=0$.} is positive definite. To do so, we will use proof by induction and
Sylvester's criterion which states that a square matrix $M$ is positive
definite if and only it has a positive determinant and all the upper-left
submatrices also have a positive determinant.

\subsection*{Proof by Induction}

Suppose that the $N\times N$ matrix $M_{N}$ whose components are
given by $A_{n,m}$ is positive-definite. Then consider the block
matrix constructed as
\begin{align}
M_{N+1}=A_{N+1,N+1}\left(\begin{array}{ccc}
M_{N} & B\\
B^{T} & 1
\end{array}\right)
\end{align}
where $B$ is a $N$-column vector with entries given by $A_{i,N+1}$.
Since $M_{N}$ is positive-definite, it has a positive determinant
and all the upper-left submatrices of $M_{N}$ also have a positive
determinant by Sylvester's criterion. To show that $M_{N+1}$ is positive-definite,
we need only show it has a positive determinant since all the upper-left
submatrices are already known .

The determinant of $M_{N+1}$ can be evaluated using the formula
\begin{align}
\det(M_{N+1})=A_{N+1,N+1}\left[2\,\det(M_{N})-\det(M_{N}+B^{T}B)\right]
\end{align}
so it is sufficient to show
\begin{align}
\det(M_{N}+B^{T}B)<2\det(M_{N})\,.
\end{align}

We denote the eigenvalues of $M_{N}+B^{T}B$ by $\lambda_{i}^{M+B}$where
they are ordered from largest to smallest $\lambda_{1}^{M+B}\geq\lambda_{2}^{M+B}\geq...\geq\lambda_{N}^{M+B}.$
Since $B^{T}B$ is a rank-one matrix, the sole non-zero eigenvalue
is given by $\beta=\Tr(B^{T}B)=\sum_{i=1}^{N}A_{i,N+1}\geq0$. Since
$B^{T}B$ is positive semi-definite, there exists an upper bound on
$\det(M_{N}+B^{T}B)$ given by the Weyl inequality $\lambda_{i}^{M+B}\leq\lambda_{i}^{M}+\beta_{i}$where
$\lambda_{i}^{M}$ are the eigenvalues of $M_{N}$ in order from largest
to smallest $\lambda_{1}\geq\lambda_{2}\geq...\geq\lambda_{N}$. We
then expand the determinant as

\begin{align}
\det(M_{N}+B^TB) & =\prod_{i=1}^{N}\lambda_{i}^{M+B} \leq\frac{\lambda_{1}^{M+B}}{\lambda^{M}}\prod_{i=1}^{N}\lambda_{i}^{M} =\left(1+\frac{\beta}{\lambda_{1}^{M}}\right)\det(M_{N})\,.
\end{align}
So it remains to show that $\lambda_{1}^{M}-\beta_{B}\geq0$ to complete
the proof. The maximum eigenvalue $\lambda_{1}^{M}$ is bound from
below by the minimum sum of a column of $M_{N}$ through the Perron-Frobenius
theorem (equivalently Gershgorin circle theorem). For the matrix $M_{N}$,
the minimum sum of a column vector is simply the sum of the $N$-th column
$\sum_{i=1}^{N}A_{i,N}$ since $A_{i,j}$ decreases with $i$ and $j$. Therefore it
remains to show
\begin{align}
\sum_{i=1}^{N}\left(A_{i,N}-A_{i,N+1}^{2}\right) \geq 0\,.
\end{align}

We split this sum up into two cases. The first case is if $N$ is
even. Then we have
\begin{align}
 & \sum_{i=1}^{N/2}A_{2i,N}-\sum_{i=1}^{(N+1)/2}A_{2i-1,N+1}^{2}=\sum_{i=1}^{n/2}\left(A_{2i,N}-A_{2i-1,N+1}^{2}\right)\label{eq:even-term}
\end{align}
since the final term in $\sum_{i=1}^{(N-1)/2}A_{2i-1,N+1}^{2}$ is
zero. Explicitly analyzing the coefficients, we see that $\left(A_{2i,N}-A_{2i-1,N+1}^{2}\right)$
is always positive for all $i\in\{1..N/2\}$, so clearly the entire
sum is positive. In the case of odd $N$, the sum becomes
\begin{align}
 & \sum_{i=1}^{(N+1)/2}A_{2i-1,N}-\sum_{i=1}^{N/2}A_{2i,N+1}^{2} =A_{N,N}+\sum_{i=1}^{N/2}\left(A_{2i-1,N}-A_{2i,N+1}^{2}\right) \,.\label{eq:odd-term}
\end{align}
Each term in this sum is also positive, so we have
shown $\lambda_{1}^{M}-\beta_{B}\geq0$. The expressions in (\ref{eq:even-term})
and (\ref{eq:odd-term}) are not obviously positive, but they reduce
to some polynomial equations which can be shown
to be positive. Therefore we've shown $\det(M_{N}+B^{T}B)<2\det(M_{N})$, thus $M_{N+1}$ is positive-definite
given that $M_{N}$ is. Since $M_{1}$ is positive-definite we completed the proof by induction. The kernel for canonical
energy is explicitly positive-semidefinite as required by the positivity of relative entropy.


\section{Rindler reconstruction for scalar operators in CFT$_2$}
\label{app:Rindler_reconstruct}
In this appendix we find an expression for the matter contribution to the second-order perturbation to the entanglement entropy of a ball $B$ using Rindler reconstruction so as to only use the one-point functions of the scalar operator in the domain of dependence $D_B$. We specialize to two dimensional CFTs in order to obtain a more explicit expression which can be compared to the gravitational contribution \eqref{result2D}. Further discussions of Rindler reconstruction can be found in the literature~\cite{Hamilton:2006az, Hamilton:2006fh, Papadodimas:2012aq, Morrison:2014jha, Almheiri:2014lwa}.

Coordinates on the Rindler wedge $R_B$ of radius $R$ can be given by $(r,\tau,\phi)$
which map back into Poincar\'e coordinates by
\begin{align}
z =& \frac{R }{r \cosh \phi + \sqrt{r^2 - 1} \cosh \tau }\,, \\
t =& \frac{R \sqrt{r^2 - 1} \sinh \tau }{r \cosh \phi + \sqrt{r^2 - 1} \cosh \tau }\,,\\
x=& \frac{R r \sinh \phi}{r \cosh \phi + \sqrt{r^2 - 1} \cosh \tau } \,,
\end{align}
where $1<r<\infty$.

The scalar field dual to an operator $\mathcal O$ can be reconstructed in this Rindler wedge using \cite{Almheiri:2014lwa}
\begin{align}
\phi(r,\tau,\phi) =& \int d\omega dk\, e^{-i \omega \tau - i k \phi} f_{\omega,k}(r) \mathcal{O}_{\omega,k}\,, \\
f_{\omega,k}(r) =& r^{-\Delta}\left(1-\frac{1}{r^2}\right)^{-i\frac\omega2} {}_2F_1 \left(
\frac\Delta2 -\frac{i (\omega + k)}{2} , \frac\Delta2 +\frac{i (\omega + k)}{2}; \Delta; r^{-2}
\right)\,,
\end{align}
where $\mathcal{O}_{\omega,k}$ is the Fourier transform of the CFT expectation value of the operator
\begin{align}
\mathcal{O}_{\omega,k} =& \int d\tau d\phi\, e^{i \omega \tau + i k \phi} \langle \mathcal{O}(\tau,\phi)\rangle\,.
\end{align}
This can be expressed in terms of the operator in the original coordinates
\begin{align}
\mathcal{O}_{\omega,k} =& \int_{D_{B}} dt dx \left[ \left( R+x+t\right)^{i\frac{k+\omega}{2}} \left(R-x-t \right)^{-i\frac{k+\omega}{2}}
\right.\nn
&\qquad\qquad\left.\left(R-x+t\right)^{i\frac{\omega-k}{2}} \left(R+x-t\right)^{i\frac{k-\omega}{2}} \right]\langle\mathcal{O}(t,x)\rangle\,,
\end{align}
where the region of integration is only over the domain of dependence $D_B$.

This form of the scalar field can be combined with \eqref{StwoM} to obtain an an expression for $\delta^{(2)}S^{scalar}$ which only depends on the expectation value of $\mathcal O$ in $D_B$,
\begin{align}
\delta^{(2)} S^{scalar}=&
-\frac14 \int_{1}^\infty dr dk d\omega_1 d\omega_2~r
\sqrt{r^2-1}  \left[ f_{\omega_1,k}(r) f_{\omega_2,-k}(r) \left( - \frac{\omega_1 \omega_2}{r^2-1} + \frac{k^2}{r^2} +\Delta(\Delta-2)  \right)
\right. \nn &\qquad\qquad+\left.
 \left(r^2-1\right) f'_{\omega_1,k}(r) f'_{\omega_2,-k}(r) \right]
 \mathcal{O}_{\omega_1,k} \mathcal{O}_{\omega_2,-k} \,.
\end{align}

\providecommand{\href}[2]{#2}\begingroup\raggedright\endgroup

\end{document}